\documentclass[a4paper]{article}

\usepackage{ISCSLP_v2}
\usepackage{bm}
\usepackage{multirow}
\usepackage[labelformat=simple]{subcaption}

\usepackage[justification=centering]{caption}
\usepackage{scalerel}

\makeatletter
\newcommand\footnoteref[1]{\protected@xdef\@thefnmark{\ref{#1}}\@footnotemark}
\makeatother

\DeclareMathSizes{10}{10}{3}{3}

\title{Voice Conversion Based on Cross-Domain Features Using Variational Auto Encoders}
\name{Wen-Chin Huang$^{12}$, Hsin-Te Hwang$^1$, Yu-Huai Peng$^1$, Yu Tsao$^3$, Hsin-Min Wang$^1$}
\address{
	$^1$Institute of Information Science, Academia Sinica, Taipei\\
	$^2$Department of Computer Science and Information Engineering, National Taiwan University, Taipei\\
	$^3$Research Center for Information Technology Innovation, Academia Sinica, Taipei}
\email{\{unilight,whm\}@iis.sinica.edu.tw; yu.tsao@citi.sinica.edu.tw}

\begin{document}

\setlength\abovedisplayskip{6pt}
\setlength\belowdisplayskip{6pt}
\setlength{\abovedisplayshortskip}{5pt}
\setlength{\belowdisplayshortskip}{6pt}

\maketitle
\begin{abstract}
An effective approach to non-parallel voice conversion (VC) is to utilize deep neural networks (DNNs), specifically variational auto encoders (VAEs), to model the latent structure of speech in an unsupervised manner. A previous study has confirmed the effectiveness of VAE using the STRAIGHT spectra for VC. However, VAE using other types of spectral features such as mel-cepstral coefficients (MCCs), which are related to human perception and have been widely used in VC, have not been properly investigated. Instead of using one specific type of spectral feature, it is expected that VAE may benefit from using multiple types of spectral features simultaneously, thereby improving the capability of VAE for VC. To this end, we propose a novel VAE framework (called cross-domain VAE, CDVAE) for VC. Specifically, the proposed framework utilizes both STRAIGHT spectra and MCCs by explicitly regularizing multiple objectives in order to constrain the behavior of the learned encoder and decoder. Experimental results demonstrate that the proposed CDVAE framework outperforms the conventional VAE framework in terms of subjective tests.
\end{abstract}

\section{Introduction}

Voice conversion (VC) aims to convert the speech from a source to that of a target without changing the linguistic content. While there are a wide variety of types and applications of VC, here we consider the most typical one, i.e., speaker voice conversion \cite{661472}. By formulating the task into a regression problem in machine learning, a conversion function that maps the acoustic features of a source speaker to those of a target speaker is to be learned. Numerous approaches have been proposed, such as Gaussian mixture model (GMM)-based methods \cite{661472,4317579}, deep neural network (DNN)-based methods \cite{5445041,6891242}, and exemplar-based methods \cite{6424242,6843941,Wu+2016}. Most of them require parallel training data, i.e., the source and target speakers utter the same transcripts for training. Since such data is hard to collect, non-parallel training has long remained one of the ultimate goals in VC.

DNNs have demonstrated its great capability in solving complex tasks in recent years, due to the rising accessibility of powerful computational resources. Recently, variational auto encoders (VAEs) \cite{2013arXiv1312.6114K} have been successfully applied to non-parallel VC \cite{7820786}. Specifically, the conversion function is composed by an encoder-decoder pair. The encoder first encodes the input into a latent content code. Then, the decoder mixes the latent content code and the target speaker code to generate the output. The encoder-decoder network and speaker codes are trained through back-propagation of the reconstruction error, along with a Kullback-Leibler (KL)-divergence loss that regularizes the distribution of the latent variable. Therefore, there is no need for parallel training data. On the other hand, cycle-consistent adversarial networks (CycleGAN) \cite{CycleGAN2017} have also been introduced to non-parallel VC \cite{2017arXiv171111293K,2018arXiv180400425F}. There are also methods that require external resources, such as transcriptions of training data, text-to-speech (TTS) systems, and speaker-independent automatic speech recognition (SI-ASR) systems. In non-parallel VC based on TTS \cite{wthkt18}, the TTS reference voices are used to create two parallel training copora: one between the source and TTS
voices and the other between the TTS and target voices. While in non-parallel VC based on SI-ASR \cite{7552917}, the phonetic PosteriorGrams (PPGs) are used to bridge the source and target voices. In this paper, we focus on VAE-based VC.

Although the effectiveness of VAE using the STRAIGHT spectra \cite{KAWAHARA1999187} for VC has been confirmed in \cite{7820786},  VAE using other types of spectral features such as mel-cepstral coefficients (MCCs) \cite{225953}, which are related to human perception and have been widely used in VC, have not been properly investigated. We expect that VAE-based VC may benefit from using multiple types of spectral features simultaneously. To this end, we propose a novel VAE framework, called cross-domain VAE (CDVAE), by extending the conventional VAE framework to jointly consider two kinds of spectral features, namely the STRAIGHT spectra (called SP for short hereafter) and MCCs. In the VAE framework for VC, an ideal, well-trained encoder is analogous to a speech/phone recognizer, such that the latent representations encoded from SP and MCCs should be similar and capable of self- or cross-reconstructing both kinds of spectral features. To achieve this goal, we introduce two additional cross-domain reconstruction errors, along with a latent similarity constraint, into the training objective.

The remainder of this paper is organized as follows. In Section~\ref{sec:sec2}, we first review the non-parallel VAE-based VC. The proposed CDVAE framework is described in Section~\ref{sec:sec3}. Experimental settings and results are presented in Section~\ref{sec:sec4}. Finally, we conclude the paper with discussions in Section~\ref{sec:sec5}. 

\section{Non-parallel Voice Conversion via Variational Auto Encoder}
\label{sec:sec2}

\iffalse
We first give a formal formulation of VC. Given a source speaker's spectral frames $\mathbf{X}_s=\{\vec{x}_{s,1},\dots,\vec{x}_{s,N}\}$ and a target speaker's spectral frames $\mathbf{X}_t=\{\vec{x}_{t,1},\dots,\vec{x}_{t,M}\}$ with $N$ and $M$ frames respectively, the goal is to find a conversion function $f$ such that
\begin{equation}
	\hat{\vec{x}}_{t,m}=f(\vec{x}_{s,n}).
\end{equation}
\fi

%\subsection{Training}

\begin{figure}[t]
	\centering
	\begin{subfigure}[b]{0.33\textwidth}
		\centering
		\includegraphics[width=0.9\textwidth]{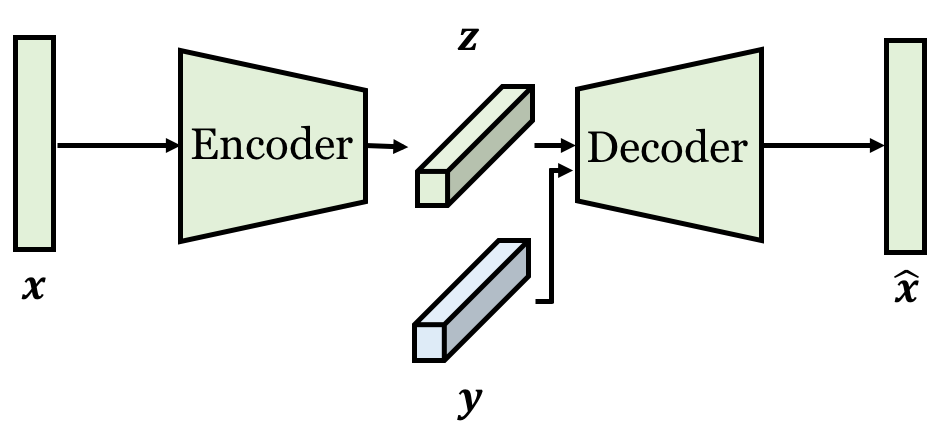} 
		\caption{VAE-based VC\label{fig:vae_model}}

	\end{subfigure}\\
	%	\vspace{0.7cm}
	\begin{subfigure}[b]{0.5\textwidth}
		\centering
		\includegraphics[width=0.9\textwidth]{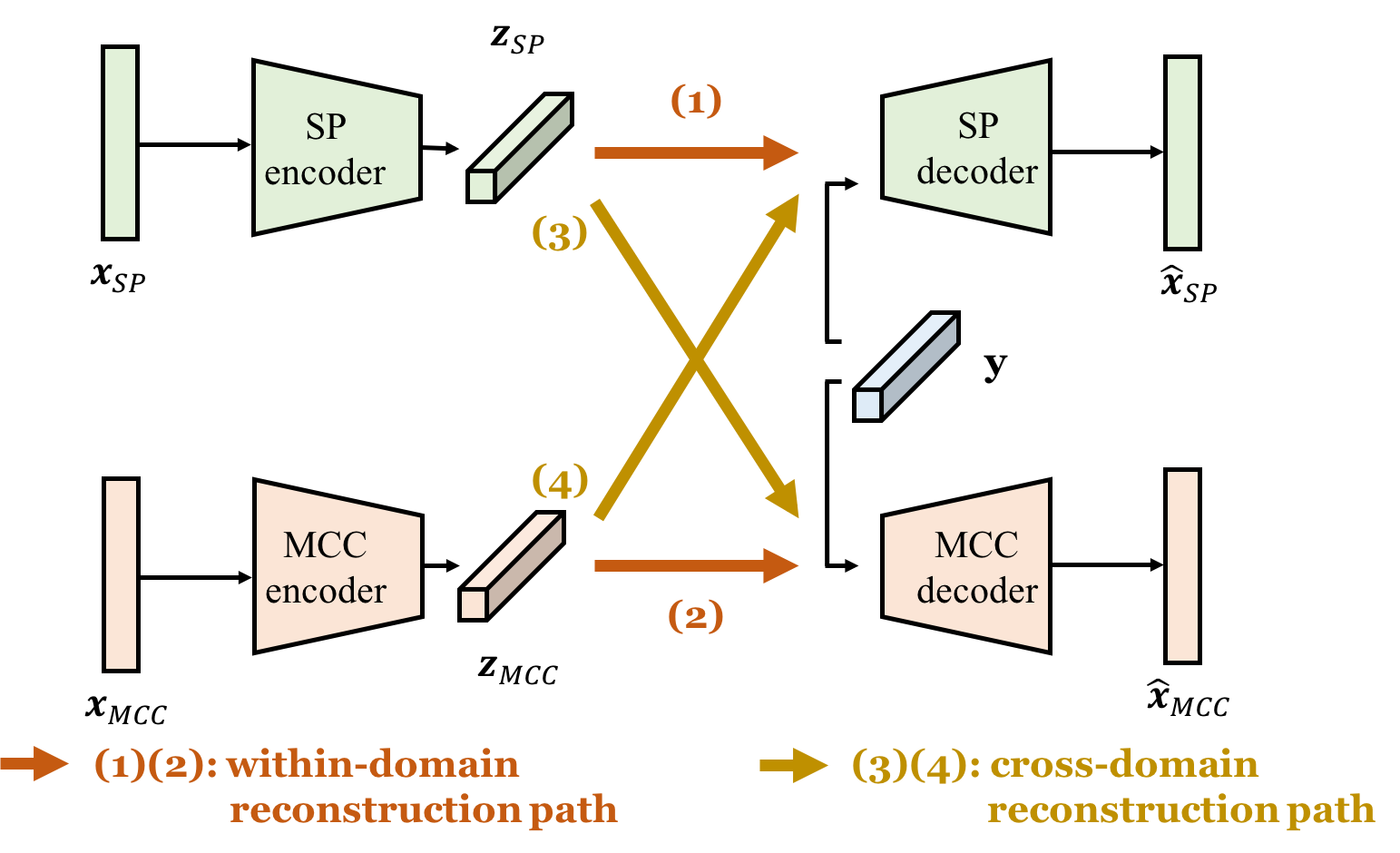}
		\caption{CDVAE-based VC\label{fig:cdvae_model}}

	\end{subfigure}
	\centering
	\caption{Illustration of the VAE-based VC and CDVAE-based VC.	\label{fig:model_structure}}
\end{figure}

Figure~\ref{fig:vae_model} depicts the structure of a typical VAE-based VC system \cite{7820786}. The conversion function is formulated as an encoder-decoder network. Specifically, Given an observed (source or target) spectral frame $\vec{x}$, a speaker-independent encoder $E_\theta$ with parameter set $\theta$ encodes $\vec{x}$ into a latent code: $\vec{z}=E_\theta(\vec{x})$. A speaker code $\vec{y}$ is then concatenated with the latent code, and passed to a conditional decoder $G_\phi$ with parameter set $\phi$ to reconstruct the input. Thus, the conversion function $f$ of VAE-based VC can be expressed as: $\hat{\vec{x}}=f(\vec{x})=G_\phi(\vec{z},\vec{y})$. The model parameters can be obtained by maximizing the variational lower bound:
\begin{equation}
	\mathcal{L}(\theta,\phi;\vec{x}, \vec{y}) = \mathcal{L}_{recon}(\vec{x},\vec{y})+\mathcal{L}_{lat}(\vec{x}),
\end{equation}
\begin{equation}
\mathcal{L}_{recon}(\vec{x},\vec{y})=\mathbb{E}_{q_\theta(\vec{z}|\vec{x})}\bigl[\log p_\phi(\vec{x}|\vec{z},\vec{y})\bigr],
\label{recon_loss}
\end{equation}
\begin{equation}
\mathcal{L}_{lat}(\vec{x})=-D_{KL}(q_\theta(\vec{z}|\vec{x}) \Vert p(\vec{z})),
\label{lat_loss}
\end{equation}
where $q_\theta(\vec{z}|\vec{x})$ is the approximate posterior, $p_\phi(\vec{x}|\vec{z},\vec{y})$ is the data likelihood, and $p(\vec{z})$ is the prior distribution of the latent space. $\mathcal{L}_{recon}$ is simply a reconstruction term as in any vanilla auto encoder, whereas $\mathcal{L}_{lat}$ regularizes the encoder to align the approximate posterior with the prior distribution.

The VAE framework makes several assumptions. First, $p_\phi(\vec{x}|\vec{z},\vec{y})$ is assumed to follow a normal distribution whose covariance is an identity matrix. Second, $p(\vec{z})$ is set to be a standard normal distribution. Third, the expectation over $\vec{z}$ is approximated by sampling via a linear-transformation based re-parameterization trick \cite{kingma2014method}. With these simplifications, we can avoid intractability and optimize the auto-encoder parameter sets $\theta\cup\phi$ via back-propagation. Note that the speaker codes can be either fixed one-hot representations \cite{7820786, Hsu2017} or learned during training (with the speaker codes randomly initialized) as in the implementation codes\footnote{\label{hsu_github_link}https://github.com/JeremyCCHsu/vae-npvc} released by the first author of \cite{7820786}.

In the conversion phase, given an input source frame, the encoder first encodes it into a latent code. Then, the decoder blends the latent code and the target speaker code to generate the converted spectral features.

\begin{table*}[ht]
	\centering
	\captionsetup{justification=centering}
	\caption{Mean Mel-cepstral distortion [dB] of all non-silent frames from the baseline and proposed frameworks.}
	\begin{tabular}{ l l c c c c }
		\toprule	
		\multicolumn{2}{l}{Method} & SF1-TF1 & SF1-TM1 & SM1-TF1 & SM1-TM1 \\
		\midrule
		\multirow{2}{*}{Baseline} & \textbf{VAE SP-SP} & 6.35 & \textbf{6.28} & 6.46 & 6.13 \\
		& \textbf{VAE MCC-MCC} & 8.26 & 9.04 & 9.01 & 7.74 \\ 
		\midrule
		\multirow{4}{*}{Proposed} & \textbf{CDVAE SP-SP} & \textbf{6.30} & 6.33 & 6.44 & 6.13 \\
		& \textbf{CDVAE SP-MCC} & 6.36 & 6.36 & 6.49 & 6.14 \\
		& \textbf{CDVAE MCC-SP} & 6.43 & 6.40 & \textbf{6.40} & 6.14 \\
		& \textbf{CDVAE MCC-MCC} & 6.47 & 6.43 & 6.47 & \textbf{6.13} \\
		\midrule
		\multicolumn{2}{l}{Before conversion} & 8.31 & 9.09 & 9.07 & 7.77 \\
		\bottomrule
	\end{tabular}
	\label{tab:MCD}
\end{table*}

\section{Cross-Domain Variational Auto Encoder for Voice Conversion}
\label{sec:sec3}

%We now describe our proposed framework, the Cross-Domain Variational Auto Encoder (CDVAE) for VC. Figure~\ref{fig:cdvae_model} depicts the overall structure.
The goal of the proposed framework is to utilize spectral features of different properties extracted from the same observed speech frame. As dipcited in Figure~\ref{fig:cdvae_model}, the CDVAE framework is a collection of encoder-decoder pairs, one for each kind of spectral feature. Here, we consider the SP and MCCs (extracted by the STRAIGHT vocoder \cite{KAWAHARA1999187}) as two kinds of spectral features (denoted as $\vec{x}_{\scaleto{SP}{3pt}}$ and $\vec{x}_{\scaleto{MCC}{3pt}}$ ). Next, we describe the training objectives and conversion procedure.
%Here we consider the SP and MCCs extracted using STRAIGHT vocoder \cite{KAWAHARA1999187}. Given an input speech frame, we can obtain SP $\vec{x}_{\scaleto{SP}{3pt}}$ and MCCs  $\vec{x}_{MCC}$ as input spectral features for the encoders of CDVAE.

\subsection{Within-domain reconstruction paths}

In Figure~\ref{fig:cdvae_model}, paths (1) and (2) depict the within-domain reconstruction paths. Specifically, the encoders first encode the corresponding input spectral features into their respective latent representations:
\begin{equation}
\begin{aligned}
\vec{z}_{\scaleto{SP}{3pt}}=E_{\scaleto{SP}{3pt}}(\vec{x}_{\scaleto{SP}{3pt}}),
\vec{z}_{\scaleto{MCC}{3pt}}=E_{\scaleto{MCC}{3pt}}(\vec{x}_{\scaleto{MCC}{3pt}}),
\label{eq:latent}
\end{aligned}
\end{equation}
where $E_{\scaleto{SP}{3pt}}$ and $E_{\scaleto{MCC}{3pt}}$ are the encoders for SP and MCCs, respectively. Blending the speaker code with the latent code, the decoders attempt to reconstruct the input spectral features:
\begin{equation}
\begin{aligned}
\hat{\vec{x}}_{\scaleto{SP}{3pt}}=G_{\scaleto{SP}{3pt}}(\vec{z}_{\scaleto{SP}{3pt}},\vec{y}),
\hat{\vec{x}}_{\scaleto{MCC}{3pt}}=G_{\scaleto{MCC}{3pt}}(\vec{z}_{\scaleto{MCC}{3pt}},\vec{y}),
\end{aligned}
\end{equation}							
where $G_{\scaleto{SP}{3pt}}$ and $G_{\scaleto{MCC}{3pt}}$ are the decoders for SP and MCCs, respectively. The \textit{within-domain reconstruction loss} is defined as:
\begin{equation}
\begin{aligned}
\mathcal{L}_{wi}&=\mathcal{L}_{recon}(\vec{x}_{\scaleto{SP}{3pt}},\vec{y})+\mathcal{L}_{recon}(\vec{x}_{\scaleto{MCC}{3pt}},\vec{y}),
\end{aligned}
\end{equation}
where $\mathcal{L}_{recon}$ is the same as $\mathcal{L}_{recon}$ in \eqref{recon_loss}. The \textit{KL-Divergence loss} is defined as:
\begin{equation}
\mathcal{L}_{KLD}=\mathcal{L}_{lat}(\vec{x}_{\scaleto{SP}{3pt}})+\mathcal{L}_{lat}(\vec{x}_{\scaleto{MCC}{3pt}}),
\end{equation}
where $\mathcal{L}_{lat}$ is the same as $\mathcal{L}_{lat}$ in \eqref{lat_loss}. Optimizing the two loss terms, $\mathcal{L}_{wi}$ and $\mathcal{L}_{KLD}$, is realized by training two VAEs for SP and MCCs, respectively. Next, we describe how we further regularize the behavior of the proposed CDVAE model.

\subsection{Cross-domain reconstruction paths}

In Figure~\ref{fig:cdvae_model}, paths (3) and (4) depict the cross-domain reconstruction paths. Specifically, for an input frame, we take the SP latent representation $\vec{z}_{\scaleto{SP}{3pt}}$ as the input of the MCC decoder (i.e., path (3)), and take the MCC latent representation $\vec{z}_{\scaleto{MCC}{3pt}}$ as the input of the SP decoder (i.e., path (4)), where $\vec{z}_{\scaleto{SP}{3pt}}$ and $\vec{z}_{\scaleto{SMCC}{3pt}}$ are obtained in \eqref{eq:latent}. The two paths also generates two outputs:
\begin{equation}
\begin{aligned}
\hat{\vec{x}}_{\scaleto{MCC}{3pt}}=G_{\scaleto{MCC}{3pt}}(\vec{z}_{\scaleto{SP}{3pt}},\vec{y}),
\hat{\vec{x}}_{\scaleto{SP}{3pt}}=G_{\scaleto{SP}{3pt}}(\vec{z}_{\scaleto{MCC}{3pt}},\vec{y}).
\end{aligned}
\end{equation}	
Therefore, we define the \textit{cross-domain reconstruction loss} as:
\begin{equation}
\begin{aligned}
\mathcal{L}_{cross}&=\mathcal{L}_{recon}(\vec{x}_{\scaleto{SP}{3pt}},\vec{y})+\mathcal{L}_{recon}(\vec{x}_{\scaleto{MCC}{3pt}},\vec{y}).
\label{eq:l_cross}
\end{aligned}
\end{equation}
In short, we introduce two extra reconstruction streams.  By optimizing the cross-domain reconstruction loss, we enforce the SP latent code to contain enough information to reconstruct the input MCCs, and vice versa. As a result, the behavior of the encoders from both feature domains are constrained to be the same, i.e., they are expected to extract similar latent information from different types of input spectral features.

\subsection{Latent similarity loss}

The cross-domain satisfaction loss in  \eqref{eq:l_cross} implicitly guarantees the latent codes of two feature types to be close to each other. To explicitly reinforce this constraint, we add a \textit{latent similarity loss} to the training objective:
\begin{equation}
\mathcal{L}_{sim}=\Vert \vec{z}_{\scaleto{SP}{3pt}}-\vec{z}_{\scaleto{MCC}{3pt}} \Vert _1 .
\end{equation}
Our preliminary results confirmed the effectiveness of introducing this loss in improving the speech quality.

\subsection{Training and conversion procedures}
\label{sec:subsec3-4}

Overall, the training objective of the CDVAE framework combines the within-domain reconstruction loss, cross-domain reconstruction loss, KL-divergence loss, and latent similarity loss:
\begin{align}
	\mathcal{L}=\mathcal{L}_{wi}+\mathcal{L}_{KLD}+\mathcal{L}_{cross}+\mathcal{L}_{sim}.
	\label{eq:total_objective}
\end{align}

The model parameters can be learned by maximizing \eqref{eq:total_objective}. In the conversion phase, there are four conversion paths (i.e., two within-domain and two cross-domain paths). Given a source speech frame, one can use either SP or MCCs as the input spectral feature. The corresponding encoder then encodes it into the latent code. Depending on the selected output spectral feature type, one then feed the corresponding decoder with the latent code and target speaker code to generate the converted spectral feature.

\section{Experiments}
\label{sec:sec4}

\subsection{Experimental settings}

The proposed CDVAE framework was evaluated on the Voice Conversion Challenge 2018 dataset \cite{2018arXiv180404262L}, which included recordings of professional US English speakers with a sampling rate of 22050 Hz. We used a subset of speakers, including two male speakers (SM1 and TM1) and two female speakers (SF1 and TF1). The training set consisted of 81 utterances per speaker while the testing set consisted of 35 utterances per speaker. Although each speaker uttered the same sentences in the corpus, we did not deliberately divide the training set into disjoint (non-parallel) subsets for two reasons: 1) The performance of the baseline VAE-based VC framework stayed unaffected regardless of the division of the training set \cite{7820786}. Similar results were also observed for the proposed framework in our preliminary experiments. 2) The training processes of the baseline and proposed frameworks did not take advantage of the alignment  information of the corpus.

The STRAIGHT vocoder \cite{KAWAHARA1999187} was used to extract speech parameters (including 513-dimensional SP, 513-dimensional AP, and $F_0$) and reconstruct the waveform. 35-dimensional MCCs (including the 0-th coefficient for the frame power) were further extracted from the SP features. Note that the SP features were normalized as described in \cite{7820786} in both baseline and proposed frameworks. In the conversion phase, for both baseline and proposed frameworks, the energy and AP were kept unmodified, and $ F_0 $ was converted using a linear mean-variance transformation in the log-$ F_0 $ domain. 
\begin{figure*}[ht]
	\centering
	\begin{subfigure}[b]{0.22\textwidth}
		\centering
		\includegraphics[width=0.85\textwidth]{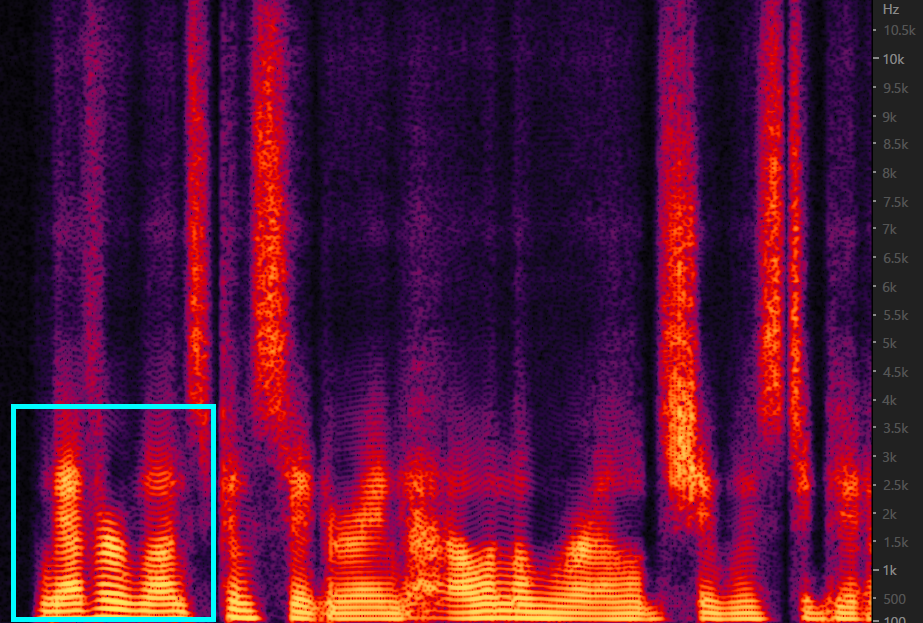}
		%\caption{\textup{\textbf{VAE SP-SP}}}
	\end{subfigure}
	\begin{subfigure}[b]{0.22\textwidth}
		\centering
		\includegraphics[width=0.85\textwidth]{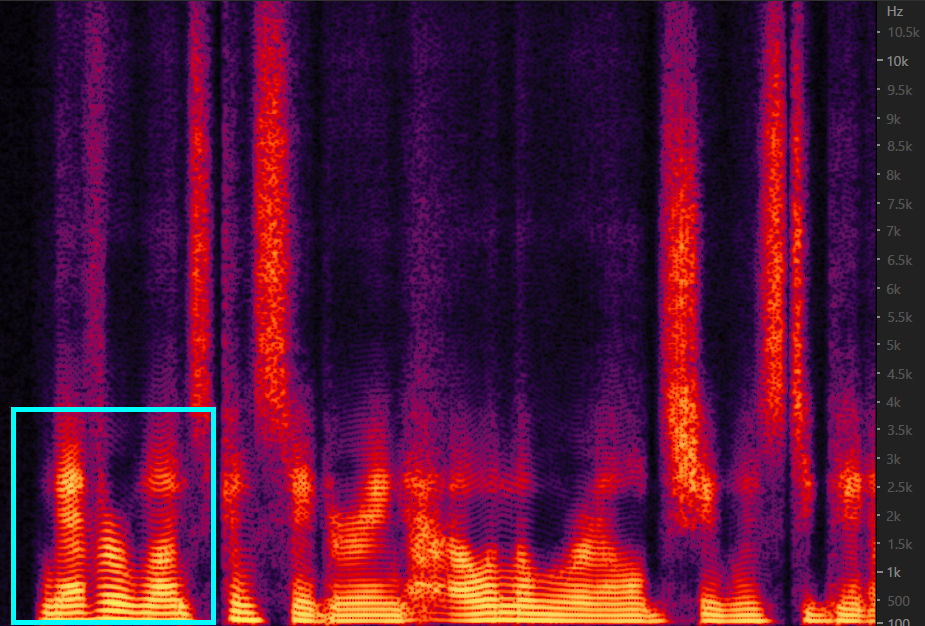}
		%\caption{\textup{\textbf{CDVAE SP-SP}}}
	\end{subfigure}
	\begin{subfigure}[b]{0.22\textwidth}
		\centering
		\includegraphics[width=0.85\textwidth]{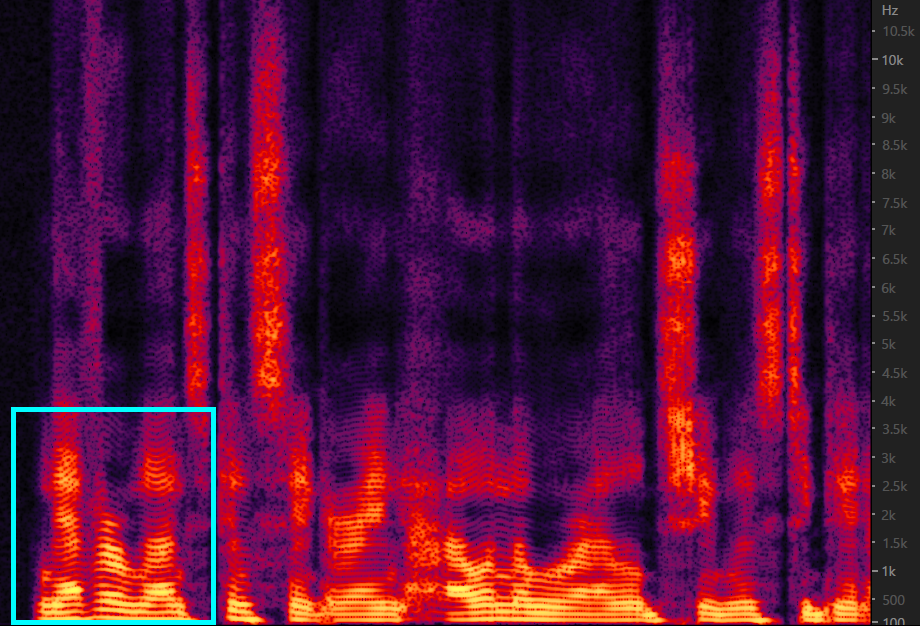}
		%\caption{\textup{\textbf{CDVAE MCC-MCC}}}
	\end{subfigure}
	
	\caption{Spectrograms of the converted speeches. Clearer formant structure by CDVAE can be observed in the blue box. From left to right: \textbf{VAE SP-SP}, \textbf{CDVAE SP-SP}, \textbf{CDVAE MCC-MCC}}.
	\label{fig:spectrograms}
\end{figure*}

\begin{figure}[t]
	\centering
	\includegraphics[width=0.25\textwidth]{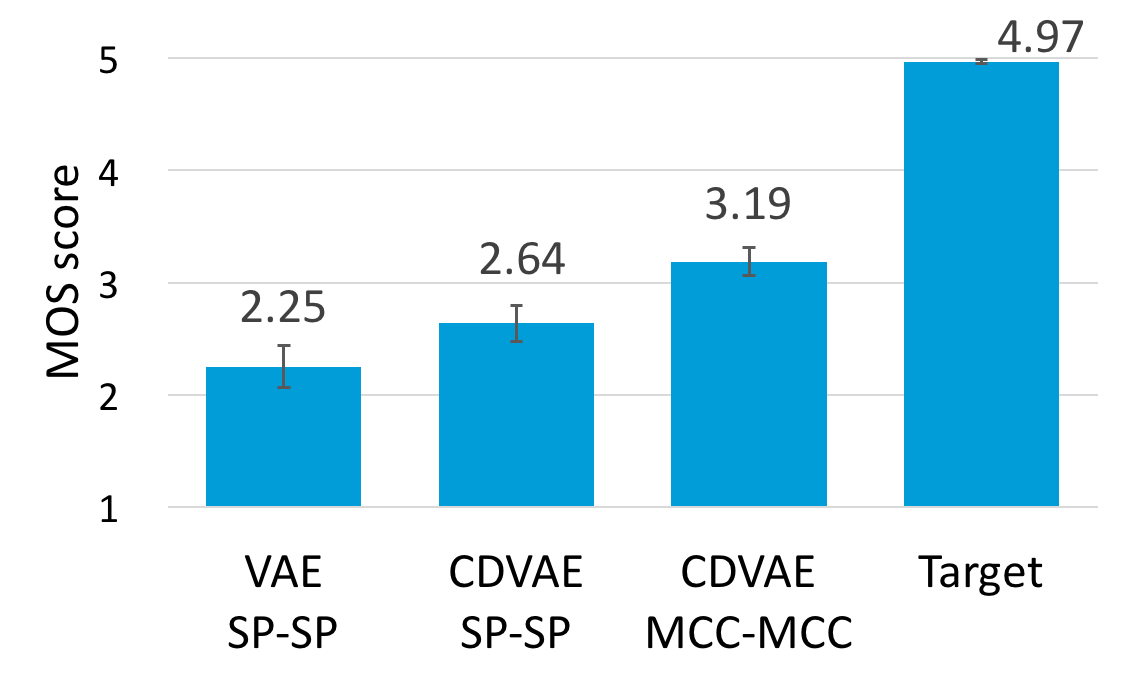}
	\caption{MOS for naturalness with 95\% confidence intervals.}
	\label{fig:Naturalness}
\end{figure}

\begin{figure}[ht]
	\centering
	\begin{subfigure}[b]{0.4\textwidth}
		\centering
		\includegraphics[width=0.85\textwidth]{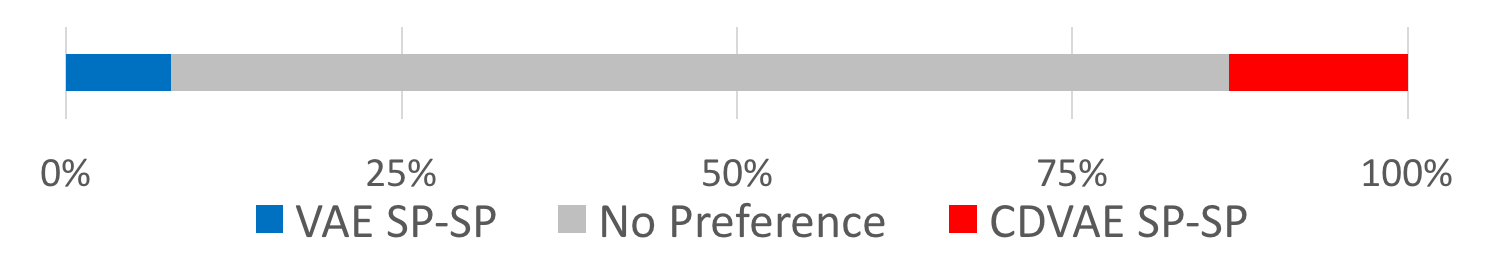}
		\caption{\textup{\textbf{VAE SP-SP}} vs. \textup{\textbf{CDVAE SP-SP}}}
	\end{subfigure}
	\begin{subfigure}[b]{0.4\textwidth}
		\centering
		\includegraphics[width=0.85\textwidth]{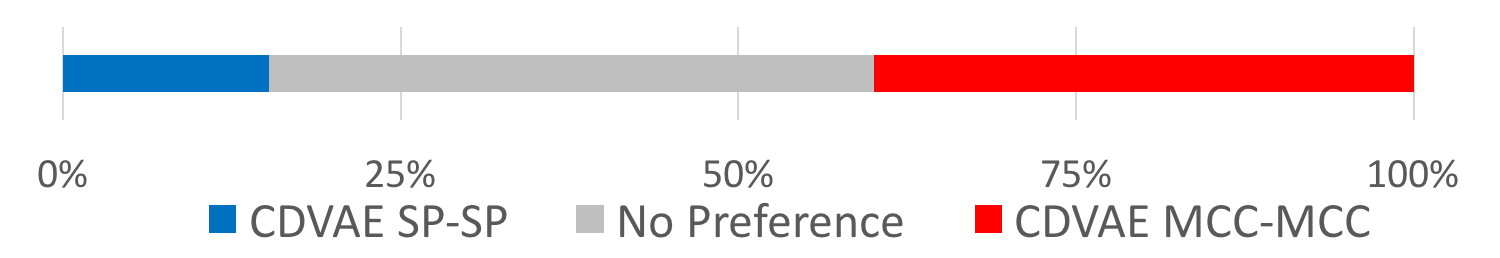}
		\caption{\textup{\textbf{CDVAE SP-SP}} vs. \textup{\textbf{CDVAE MCC-MCC}}}
	\end{subfigure}

	\caption{Preference on speaker similarity.}
	\label{fig:Similarity}
\end{figure}

\subsection{Evaluations}

We compared the proposed CDVAE framework with the baseline VAE framework \cite{7820786}. Specifically, we trained three models and evaluated their output results:
\begin{itemize}
	\setlength{\parskip}{0pt}
	\item \textbf{VAE SP-SP}: The baseline VAE framework trained on SP as described in Section~\ref{sec:sec2}.
	\item \textbf{VAE MCC-MCC}: Another baseline VAE framework as described in Section~\ref{sec:sec2}, but trained on MCCs.
	%The input and output feature domain pairs are both MCCs.  
	\item \textbf{CDVAE}: The proposed CDVAE framework as described in Section~\ref{sec:sec3}, trained on both SP and MCCs. The converted spectral features obtained from path (1) to (4) as depicted in Figure~\ref{fig:cdvae_model} were referred to \textbf{CDVAE SP-SP}, \textbf{CDVAE MCC-MCC}, \textbf{CDVAE SP-MCC}, and \textbf{CDVAE MCC-SP}, respectively.
\end{itemize}

We used Hsu's codes\footnoteref{hsu_github_link} to construct the baseline systems (i.e., \textbf{VAE SP-SP} and \textbf{VAE MCC-MCC}). Specifically, the baseline systems consisted of a CNN \cite{NIPS2012_4824} -based encoder and decoder. Layer normalization \cite{2016arXiv160706450L} was applied after each layer except for the last layer of the decoder. The latent space was 128-dimensional, and the output of the encoder contained the mean and log-variance vectors of the latent distribution. The speaker code was 128-dimensional with random initialization, simultaneously optimized with the encoders and decoders. The size of mini-batch was 16, and the optimizer was ADAM \cite{kingma2014method} with a constant 0.0001 learning rate. The proposed system simply consisted of two VAEs with the same network architectures and training hyperparameters, except that we empirically used a mini-batch of 1.

\subsubsection{Objective Evaluations}

We reported mean Mel-cepstral distortion (MCD) values on the testing set to evaluate the proposed and baseline frameworks. The results in Table~\ref{tab:MCD} show that the proposed framework successfully performed spectral conversion in both SP and MCC domains (cf. \textbf{CDVAE SP-SP} and \textbf{CDVAE MCC-MCC}) while the baseline VAE framework only performed well in the SP domain (cf. \textbf{VAE SP-SP}). The MCD values of \textbf{VAE MCC-MCC} were almost identical to those before conversion, implying that the VAE framework totally failed in the MCC domain. Similar results were also found in a recent study \cite{2945-18}. In addition, \textbf{CDVAE MCC-MCC} even outperformed the baseline \textbf{VAE SP-SP}. To our best knowledge, this is the first time that VAE-based VC successfully works on MCCs. This result demonstrates the potential of our proposed framework in various extensively studied low-dimensional perceptual features. From Table~\ref{tab:MCD}, we also observe that CDVAE-based within-domain conversion and cross-domain conversion were equally successful. The result further confirmed that the additional \textit{cross-domain reconstruction loss} and \textit{latent similarity loss} did play a good role in learning the latent representation of speech.  

Figure~\ref{fig:spectrograms} shows the spectrograms of the converted speeches obtained by \textbf{VAE SP-SP}, \textbf{CDVAE SP-SP} and \textbf{CDVAE MCC-MCC}. We can see that \textbf{CDVAE MCC-MCC} produced more spectral details, particularly in the higher frequency bands (4k-8kHz). Clearer formant structures in the lower frequency bands can also be observed in our CDVAE framework, particularly \textbf{CDVAE MCC-MCC}, as highlighted in the figure.

\subsubsection{Subjective Evaluations}

We chose a subset of systems for subjective evaluation, namely \textbf{VAE SP-SP}, \textbf{CDVAE SP-SP}, and \textbf{CDVAE MCC-MCC}. The \textbf{VAE MCC-MCC} system was eliminated because of poor performance. \textbf{CDVAE SP-MCC} and \textbf{CDVAE MCC-SP} were eliminated because they were considered auxiliary by-products and the naturalness and speaker similarity of the output speeches did not stand out from others.

For each conversion pair, ten sentences were randomly selected from the testing set, thereby resulting in 40 (4$\times$10) test sentences. Nine subjects were recruited to conduct the naturalness and speaker similarity tests. For all the compared systems, the global variance post-filtering method \cite{Silén12waysto} and a low-pass filter with a Gaussian window were applied to the converted spectral features to overcome the discontinuity and over-smoothing problems.

First, we conducted the mean opinion score (MOS) test using a five-point scale for naturalness evaluation. Figure~\ref{fig:Naturalness} depicts the overall average scores (including the score of natural target speech). The results demonstrate that two proposed systems outperformed the baseline system, and \textbf{CDVAE MCC-MCC} outperformed \textbf{CDVAE SP-SP}. This is encouraging, since our initial motivation is to improve naturalness using perception-based spectral features such as MCCs instead of SP.

Next, we conducted the ABX test to evaluate speaker similarity as described in \cite{7820786}. The results in Figure~\ref{fig:Similarity} show that the proposed \textbf{CDVAE SP-SP} system slightly outperformed the baseline \textbf{VAE SP-SP} system, and \textbf{CDVAE MCC-MCC} was superior to \textbf{CDVAE SP-SP}. Overall, our systems outperformed the baseline system in both subjective tests.\footnote{\label{demo_link}https://unilight.github.io/CDVAE-Demo/}

\section{Discussion}
\label{sec:sec5}

In Section~\ref{sec:sec4}, we have shown that our proposed CDVAE framework successfully utilizes cross-domain features to improve the capability of
VAE for VC, and outperforms the baseline VAE-based VC system in the subjective tests. The question is: how does the underlying speech model benefit from our framework?

Recall that the viability of the VAE framework relies on the decomposition of input frames, which is assumed to be composed of a latent code (in VC, phonetic code or linguistic content) and a speaker code. Ideally, when applying VAE to VC, the latent code should contain solely the phonetic information of the frame, with no information about the speaker. However, this decomposition is not explicitly guaranteed. Hand-crafted features like SP or MCCs possess their own natures, thus even for the same input frame, the required information to reconstruct the inputs from different feature domains may differ.  When trained with one feature alone, only the necessary information to reconstruct that feature is left in the latent code, thus the VAE framework might fit the property of that specific feature too well, losing the generalization ability. One way to reinforce decomposition is to involve as many speakers as possible during training, which may not necessarily lead to better decomposition. Our proposed framework forces the encoder to act more like a speaker-independent phone recognizer, thus filters out unnecessary, speaker-dependent information of the input feature. As a result, our framework not only achieves cross-domain feature property satisfaction, but learns more disentangled latent representation of speech.
 
In the future, we plan to investigate in detail the above assumption. In addition, Wasserstein generative adversarial network (WGAN) \cite{2017arXiv170107875A} has been introduced to the conventional VAE-based VC method \cite{Hsu2017} for improving the naturalness of converted speech, so we also plan to introduce WGAN to the proposed framework.

\section{Acknowledgement}
This work was supported in part by the Ministry of Science and Technology of Taiwan under Grants: MOST 105-2221-E001-012-MY3 and MOST 107-2221-E-001-008-MY3.

\bibliographystyle{IEEEtran}

\bibliography{mybib}

\end{document}